# High-T$_c$ superconductivity in entirely end-bonded multi-walled carbon nanotubes

I.Takesue[1,4], J.Haruyama[1,4,*], N.Kobayashi[1], S.Chiashi[2], S.Maruyama[2], T.Sugai[3,4], H.Shinohara[3,4]

[1]*Aoyama Gakuin University, 5-10-1 Fuchinobe, Sagamihara, Kanagawa 229-8558 Japan*
[2]*Tokyo University, 7-3-1 Hongo, Bunkyo-ku, Tokyo 113-0033 Japan*
[3]*Nagoya University, Furo-cho, Chigusa, Nagoya   464-8602   Japan*
[4]*JST-CREST, 4-1-8 Hon-machi, Kawaguchi, Saitama 332-0012 Japan*

* Corresponding author; J-haru@ee.aoyama.ac.jp

We report that entirely end-bonded multiwalled carbon nanotubes (MWNTs) can exhibit superconductivity with a transition temperature (T$_c$) as high as 12 K, which is approximately 30 times greater than T$_c$ reported for ropes of single-walled nanotubes. We find that the emergence of this superconductivity is very sensitive to the junction structures of the Au electrode/MWNTs. This reveals that only MWNTs with optimal numbers of electrically activated shells, which are realized by end bonding, can allow superconductivity due to intershell effects.

One-dimensional (1D) systems face some obstructions that prevent the emergence of superconductivity, such as (1) Tomonaga-Luttinger liquid (TLL) states consisting of repulsive electron-electron (e - e) interaction [1] - [3], (2) Peierls transition (charge-density waves), and (3) a small density of states, which becomes significant when the Fermi level is not aligned with van-Hove singularities (VHSs). A carbon nanotube (CN), an ideal 1D molecular conductor, is one of the best candidates for investigating the possibility of 1D superconductivity and the interplay of 1D superconductivity with the abovementioned onstructions. A variety of intriguing quantum phenomena in CNs has been reported; however, only two groups to our knowledge have experimentally reported superconductivity (e.g., with a transition temperature (T$_c$) as low as ~0.2K in ropes of single-walled CNs (SWNTs) [4] and that was identified only from the Meissner effect in arrays of thin SWNTs (diameter ~ 0.4nm)[5]). In addition, the interplay of superconductivity with the abovementioned 1D phenomena has not been investigated.

From the viewpoint of sp$^x$- orbital related superconductivity, it was recently reported that B-doped diamond and CaC$_6$ could exhibit superconductivity [6]. These findings stress a high potentiality of carbon-related materials as superconductors and, hence, CNs are also promising for this purpose.

From theoretical standpoints, it was predicted in refs. [7]–[10] that the interplay of TLL states with superconductivity is highly sensitive to the phonon modes, electron-phonon (e-p) coupling, strength of the short-range effective attractive interaction obtained after screening the e-e interaction, and structures of CN ropes. These factors would permit the appearance of superconductivity in specific cases. In ref. [11] as well, it was predicted that the e-p interaction is important for superconductivity in very thin CNs with sp$^3$ hybrid orbitals and an out-of-plane optical phonon mode. It should be mentioned that these theoretical predictions have not yet been experimentally verified. To the best of our knowledge, confirmed reports of superconductivity in 1D conductors is only in organic materials [12].

Here, in refs. [13] - [15], we reported the successful realization of end-bonding of multi-walled CNs (MWNTs) that were synthesized in nanopores of alumina templates. Further, we recently realized proximity-induced superconductivity (PIS) by end-bonding MWNTs, which were prepared using the same method, in Nb/MWNTs/Al junctions [13, 14]. They proved that Cooper pairs could be effectively transported without destruction only through the highly transparent interface of the CNs/Nb junctions obtained by this end-bonding. Such entire end-bonding has never been carried out in conventional field-effect transistor (FET) structures using CNs as the channels.

In this study, we followed the method of using nanoporous alumina templates but employing some specific conditions [16] (Fe/Co catalyst and methanol gas) in order to synthesize arrays of Au/MWNTs/Al junctions (Fig.1(a)); then, we investigated the possibility of superconductivity in MWNTs. Figure 1(d) shows a plane transmission electron microscope (TEM) image of the array shown in (a). MWNTs are evidently visible in many pores. In the inset, a high-resolution cross-sectional TEM image of a MWNT is shown. The structure of this MWNT is similar to those in conventional MWNTs. Indeed, it was confirmed that some MWNTs includes no Fe/Co catalyst, which tends to destroy Cooper pairs, in the entire region. Figure 1(e) shows the result of resonance Raman measurements of the MWNTs. A large peak is observable significantly around 1600 cm$^{-1}$ (the so-called G-band). This strongly indicates that the MWNTs are of a high quality and do not have defects. The absence of both the ferromagnetic catalyst and defects and the high quality in the MWNTs are very different from the case of MWNTs synthesized in our previous studies [13]-[15].

In order to investigate importance of end-bonding and intershell effects for realization of superconductivity, we prepared the following three-different types of Au electrode/MWNTs junctions using this MWNT as shown in Fig.1 [17]; (1) Entire Au-end junctions (Fig.1(a)), (2) Partial Au-end junctions (Fig.1(b)), and (3) Au-bulk junctions (Fig.1(c)). Entire Au-end junctions can be realized by sufficiently cutting the MWNTs accumulated on the template surface [17]. For such junctions, it has been proved that this method can allow making contact of Au electrode to the entire circumference of the top end of one shell and making such a contact to all the shells of a MWNT [13]-[15] (red lines in Fig.1(a)), resulting in the entire end-bonding. In contrast, in the case of partial Au-end junctions that are obtained from insufficient cutting, only partial shells can have end contacts to the Au electrode (Fig.1(b)). Finally, for Au-bulk junctions that are obtained without cutting, only the outermost shell can contact with the Au electrode, as reported in previous studies [18, 19] (Fig.1(c)). Each structure was confirmed by high-resolution cross-sectional TEM observations.

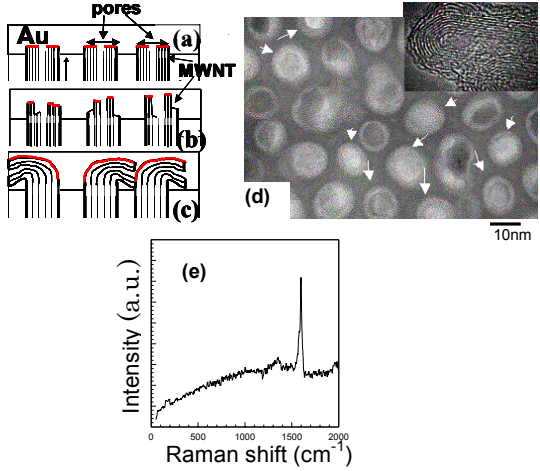
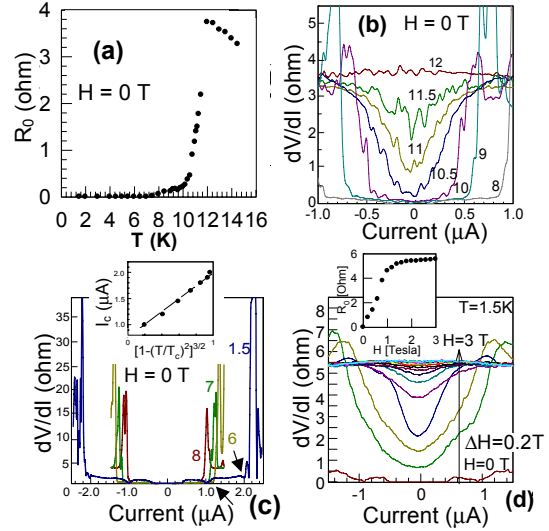

**Fig. 1:** Schematic cross-sections of Au/MWNTs interfaces in Au/MWNTs/Al junctions prepared in nano-pores of alumina templates [13 – 15] ]: **(a)** Entire Au-end, **(b)** Partial Au-end, and **(c)** Au-bulk junctions. Individual nano-pore includes one MWNT. The solid lines in a MWNT denote the shells and the red lines denote the shells where electrical contacts to Au electrodes can be present. The lengths of the MWNTs are ~0.6 µm. Quasi-four-terminal measurements were performed by attaching sets of two lead lines to the Au electrode and Al substrate. Approximately $10^3$~$10^4$ MWNTs exist under an electrode. **(d)** A plane TEM image of the MWNT array, which was observed around the top end of Fig.1(a). MWNTs were indicated by arrows. Inset: High-resolution cross-sectional TEM image of a typical MWNT. The MWNT has an outer diameter of ~7.4 nm, inner diameter of ~2 nm, shell thickness of ~2.7 nm, and nine shells. **(e)** The result of resonance Raman measurements of the MWNT at a laser energy of 2.41 eV.

**Fig. 2:** Electrical properties of the sample with the entire Au-end junction. A residual resistance of <~1 Ω has been subtracted in the figures [20]. **(a)** The zero-bias resistance ($R_0$) as a function of the temperature at zero magnetic field (H = 0 T). **(b)(c)** The differential resistance as a function of the current for different temperatures > $T_c$ (b) and < $T_c$ (c). The numbers on each curve denote the temperatures in Kelvin. The critical currents ($I_c$) were identified in contradiction to [4], although the slight resistance increases due to the residual resistances exist in the large current regions even in a superconductive state (c). We defined the current, at which an abrupt resistance increase appears, as $I_c$. The presence of resistance oscillations in the larger current regions is consistent with [4, 13] (interpreted as a phase slip). Inset of (c): $I_c$, which corresponds to the current values indicated by arrows in the main panel, as a function of the temperature. $I_c$ was normalized for the Ginzburg-Landau behavior of $I_c$. The solid line is provided just to eyes. **(d)** The differential resistance as a function of the current for different magnetic fields. The number on each curve denotes the magnetic field (in Tesla) that was applied perpendicular to the tube axis. Inset: $R_0$ as a function of the magnetic field.

Figure 2 shows the electrical properties of the sample with the entire Au-end junction. Figure 2(a) shows the zero-bias resistance ($R_0$) as a function of the temperature. $R_0$ increases with decreasing temperature and the following can be evidently confirmed: an abrupt superconducting transition with the onset $T_c$ as high as 12 K exists, and the temperature where $R_0$ drops to 0 Ω ($T_c$ (R = 0)) is as high as 7.8 K [20]. These values for the onset $T_c$ and $T_c$ (R = 0) are at least about 30 and 40 times greater, respectively, than those reported for SWNT ropes [4].

Figure 2(b) shows the differential resistance as a function of the current for different temperatures. A low and broad resistance peak exists at T = 12K. This peak disappears suddenly and a resistance dip appears at T = 11.5K. The depth and width of this resistance dip monotonically increase as the temperature decreases, corresponding to the abrupt $R_0$ drop in Fig.2(a), and attain at 0 Ω at T = ~8K. The value of the superconducting gap Δ ≈ 1.15 meV, which was estimated from the dip in the differential resistance as a function of the voltage at T = 8K in this sample, is in excellent agreement with the Bardeen-Cooper-Schrieffer (BCS) relation Δ = 1.76k$T_c$, when $T_c$(R=0) = 7.8K is employed. Moreover, the behavior of the critical current ($I_c$) below $T_c$, which is shown in Fig.2(c), as a function of the normalized temperature (inset of Fig.2(c)) is also in excellent qualitative agreement with the Ginzburg-Landau (GL) critical current behavior for a homogeneous order parameter, $I_c \propto [1- (T/T_c)^2]^{3/2}$ [21]. These agreements support that the abrupt $R_0$ drop observed in Fig.2(a) and the corresponding dip in the differential resistance in Fig.2(b) and (c) are indeed attributed to superconductivity, which is strongly related to BCS-type superconductivity.

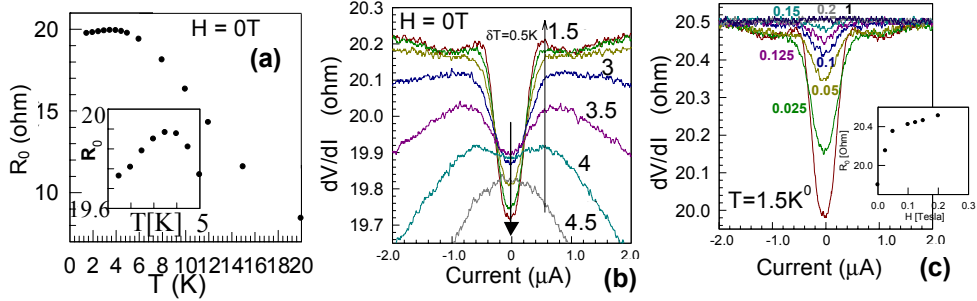

**Fig. 3:** Electrical properties of the sample with the partial Au-end junction. **(a)** $R_0$ as a function of the temperature. Inset: Enlarged view of the main panel for $1.5K \leq T \leq 5K$. **(b)** Differential resistance as a function of the current for different temperatures. **(c)** Differential resistance as a function of the current for different magnetic fields. Inset: $R_0$ as a function of the magnetic field.

Figure 2(d) shows the differential resistance as a function of the current for different magnetic fields (H). The resistance dip actually disappears due to the applied fields—similar to the conventional superconducting behavior—as shown in the main panel of Fig.2(d). The drastic increase in $R_0$ as the field increases from zero (inset) differs greatly from the behavior of $R_0$ in SWNT ropes [4]. This low critical field and the estimated magnetic penetration length > 10nm stress that the observed superconductivity is type-II without defects for pinning of the magnetic fluxes.

Consequently, we have confirmed that entirely end-bonded MWNTs exhibit superconductivity with $T_c$ as high as 12 K, from the following facts: (1) $R_0$ drops abruptly to 0 Ω, (2) the width of the differential resistance dip is in agreement with the superconducting gap $\Delta$ in accordance with the BCS relation $\Delta = 1.76kT_c$, (3) the critical current behavior qualitatively follows the relation $I_c \propto [1 - (T/T_c)^2]^{3/2}$, and (4) the dip in the differential resistance disappears due to the applied magnetic field. We confirmed such high-$T_c$ characteristics (i.e., onset $T_c = 6$~12 K) only in six samples to date.

In addition, we have found that the samples with $T_c < 4K$ (i.e., Fig.3-samples) have indeed exhibited diamagnetism. However, this diamagnetism appeared at about $T = 30$ ~ 40K and gradually dropped with decreasing temperatures through any samples. Furthermore, no abrupt change in the magnetization was observed at $T < 4K$. Hence, we conclude that this diamagnetism cannot not be attributed to the Meissner effect. Thus, the Meissner effect has not yet been found to date. We believe that very careful treatment is required for identifying the Meissner effect in CNs. For instance, it may be difficult to understand how the screening current (loop current) that leads to the Meissner effect can flow through a rope of SWNTs or a MWNT, because one rope (MWNT) includes many SWNTs (shells) located at random with different chiralities and diameters. The strength of the intertube Josephson coupling (intertube resistance) is also not uniform. From this viewpoint, the present study is analogous to ref.[4], which reported superconductivity but did not mention the Meissner effect [27].

In contrast, neither a $R_0$ drop nor a dip in the differential resistance was found in the measured temperature range in any Au-bulk junction samples.

As shown in the main panel of Fig.3(a), in the partially end-bonded sample, $R_0$ increases with decreasing temperature and gradually saturates. Then, only a small $R_0$ drop (i.e., a sign of superconductivity) appears below $T=$ ~3.5K without a decrease in $R_0$ down to 0 Ω at $T = 1.5K$ (inset of Fig.3(a)). Most of the samples have shown this sign of superconductivity.

This behavior can be easily understood by observing the behavior of the differential resistance as a function of the current for different temperatures as shown in Fig.3(b), which differs greatly from the results in Fig.2(b). A large and broad resistance peak is observable at $T = 4.5K$. It grows and broadens as the temperature decreases, corresponding to the $R_0$ increase in Fig.3(a), and almost completely disappears at $T = 2.5K$. In contrast, a resistance dip with a narrow width begins to appear at the center of this peak at $T = 4$ K and its depth monotonically increases as the temperature decreases.

Consequently, a corresponding small drop in $R_0$ can appear only below $T = 3.5$ K (Fig. 3(a)), which results from the superposition of the growth of the resistance peak and the deepening of the resistance dip at zero current. Importantly, presence of this large resistance peak prevents both the emergence of the resistance dip at $T \geq 4.5$ K and the $R_0$ drop at $T \geq 3.5K$, in contrast to Fig.2. This competition between the dips and peaks in the differential resistance has also been reported in the observation of PIS in SWNTs attached to Nb electrodes [22]. Our results are qualitatively consistent with those.

The disappearance of the resistance dip and the increase in $R_0$ in Fig.3(c) are qualitatively similar to those in Fig.2(d). Because a small dip in the differential resistance is a sign of superconductivity, the critical field is extremely small in this case.

Here, we discuss the origins of the observed differences in the superconductivity, which strongly depended on the junction structures of Au/MWNTs. One of the origins is the effective transport of Cooper pairs from the MWNTs to the Au electrode via the highly transparent interface, which were realized only in the entire Au-end junctions, as proved in refs. [13] - [15].

Moreover, the power law behaviors in the relationships of $G_0$ versus temperature (i.e., $G_0 \propto T^\alpha$; monotonic increase in $R_0$ with decreasing temperature) and their interplay with the emergence of superconductivity, which strongly depends on the junction structures, were confirmed at temperatures > $T_c$ for all the samples with different junctions. These stress the possible presence of the competition between the superconductivity and TLL states as one of the origins of the observed junction dependence.

TLL states, which are a non Fermi-liquid state arising from 1D repulsive electron-electron interaction, have been frequently reported by showing power laws (i.e., $G_0 \propto E^\alpha$, where E is the energy) in both MWNTs [1] and SWNTs[2]. As mentioned in the introduction, the interplay of TLL states with superconductivity has recently attracted considerable attention [13]–[15]. The value of $\alpha$ indicates the strength of the e-e interaction and was extremely sensitive to the junction structures of electrodes/CNs. Ref. [1] reported the value of $\alpha_{bulk}$ = ~ 0.3 for an Au-bulk junction and the value of $\alpha_{end}$ = ~ 0.7 for an Au-end junction [24]. We estimated $\alpha$ = ~ 0.7, ~ 0.8, and ~ 0.3 from the power laws at T > $T_c$ for Figs.2 (a), 3(a), and the measurement result of the Au-bulk sample, respectively. These values are in good agreement with the abovementioned values of $\alpha_{end}$ and $\alpha_{bulk}$. This agreement proves the actual presence of Au-end and Au-bulk junctions in our systems as well as the presence of TLLs in these MWNTs.

Only the $G_0 \propto T^{~0.3}$ relationship was confirmed for all the temperatures in the Au-bulk junction sample. This is consistent with refs. [18, 19], which reported that only the (second) outermost shell became electrically active in the Au-bulk junction of MWNT-FETs and that exhibited TLL states in MWNTs. This explains why most of the MWNT-FETs with electrode-bulk junctions in previous studies did not exhibit superconductivity. This result means that a SWNT with a large diameter cannot take a superconducting transition because superconductivity cannot overcome TLL states [7].

In contrast, in the sample with the entire Au-end junction, the $G_0 \propto T^{-0.7}$ relationship at T > $T_c$ abruptly disappeared and a superconductive phase emerged at $T_c$ = 12 K as the temperature decreased. On the other hand, in the sample with the partial Au-end junction, the $G_0 \propto T^{-0.8}$ relationship gradually saturated and a sign of superconductivity appeared at about T = 3.5 K. The former observation reveals that superconductivity can easily overcome TLL states, while the latter implies that superconductivity competes with TLL states as discussed in [22]. In Figs.2(b) and 3(b), these difference in the competition between TLL states and the superconducting phase correspond to the difference in the competition between the peaks and the dips in the differential resistance, respectively.

The entire end bonding of MWNTs made all the shells electrically active, while only some of the shells were electrically active in the partial Au-end junctions. These indicate that the abovementioned competition between TLL states and superconductivity is at least strongly associated with the number of electrically active shells (N) of the MWNTs (i.e., N = 1 for the Au bulk sample, N = 9 in Fig.2(a), and 1 < N < 9 in Fig.3(a)). This stresses that intershell effects in the MWNTs play a key role in the emergence of high-$T_c$ superconductivity, which overcomes TLL states.

In fact, the intershell effects in MWNTs have been discussed for TLL states, predicting that TLL states are sensitive to N [1, 3] and that this theory is applicable to SWNT ropes. Furthermore, the importance of intertube effects in the appearance of the competition between superconductivity and TLL states was indeed predicted in the case of SWNT ropes [8], as described below.

TLL states were suppressed by intertube electrostatic charge coupling (i.e., coupled TLLs; sliding TLLs [23]), because the intertube single electron tunneling was prohibited due to the misalignment of carbon atoms between neighboring SWNTs with different chiralities and diameters in a rope. In contrast, intertube Cooper-pair tunneling was allowed when TLL states were suppressed and, hence, the intratube short-range effective attractive interaction obtained after screening TLL states could sufficiently grow over the SWNTs as the temperature decreased. Both these effects—and consequently $T_c$—were enhanced as the number of SWNTs in a rope and the strength of the intratube attractive interaction increased.

Our results can be qualitatively interpreted by replacing SWNTs in a rope in this model [8] with shells in a MWNT (i.e., replacing intertube effects with intershell effects). Because the differences in chiralities and diameters among shells in a MWNT are greater than those among SWNTs in a rope, this model can lead to more significant results in the case of MWNTs. For a quantitative interpretation of these intershell effects, it is crucial to experimentally clarify the dependence of $T_c$ on N, $\alpha$, and the strength of the intrashell attractive interaction.

For future research, the following investigations are required: (1) the reproducibility of the entire end-bonding and the high $T_c$ should be improved, (2) presence of the Meissner effect should be carefully confirmed, (3) $T_c$ should be increased by intentional carrier doping, (4) the influence of curvature (i.e., a thin tube structure) in MWNTs on the e-p interaction and its interplay with $T_c$ [11, 26] should be investigated in comparison with $MgB_2$, graphite-based superconductors [6], and alkali-doped fullerenes, (5) the influence of coupling of neighboring MWNTs in an array should be studied because this might screen the e-e interaction, and (6) the interplay of VHSs in MWNTs with $T_c$ should be clarified. With regard to point 4, our MWNT may include thin SWNTs (diameter << 1 nm) in the core [11]. Finally, these superconducting MWNTs are expected to be applied for molecular quantum computation [14, 19, 25]. A MWNT can be expected to be a molecular conductor, which is promising for 1D superconductivity.

We are grateful to J.Akimitsu, R.Saito, S.Saito, S.Tarucha, Y.Iye, M.Tsukada, J.Gonzalez, H.Bouchiat, E.Demler, R.Barnett, R.Egger, G.Loupias, J.-P. Leburton, D.Loss, and M.Dresselhaus for fruitful discussions and encouragement. We also thank N.Sugiyama for acquiring the nice plane TEM images and J.Mizubayashi for the assistance in obtaining the Raman measurements.